\documentclass[aps,prl,twocolumn,superscriptaddress]{revtex4}

\usepackage{graphicx}

\usepackage{amsmath}

\newcommand{\EqLabel}[1]{\label{#1}}

\begin{document}

\title{The dynamics of a doped hole in cuprates is not controlled by
 spin fluctuations} 

\author{Hadi Ebrahimnejad} 
\affiliation{Department of Physics and Astronomy, University of
British Columbia, Vancouver B.C. V6T 1Z1, Canada} 

\author{George A. Sawatzky}
\affiliation{Department of Physics and Astronomy, University of
British Columbia, Vancouver B.C. V6T 1Z1, Canada}
\affiliation{Quantum Matter Institute, University of
British Columbia, Vancouver B.C. V6T 1Z4, Canada}

\author{Mona Berciu}
\affiliation{Department of Physics and Astronomy, University of
British Columbia, Vancouver B.C. V6T 1Z1, Canada}
\affiliation{Quantum Matter Institute, University of
British Columbia, Vancouver B.C. V6T 1Z4, Canada}

\maketitle

{\bf Twenty seven years after the discovery of high-temperature
  superconductivity \cite{BedMu}, consensus on its theoretical
  explanation is still absent. To a good extent, this is due to the
  difficulty of studying strongly correlated systems near
  half-filling, needed to understand the behaviour of one or few holes
  doped into a CuO$_2$ layer. To simplify this task it is customary to
  replace three-band models \cite{Emery} describing the doping holes
  as entering the O $2p$ orbitals of these charge-transfer insulators
  \cite{ZSA} with much simpler one-band Hubbard or $tJ$ models
  \cite{rev1,rev2}. Here we challenge this approach, showing that not
  only is the dynamics of a doped hole easier to understand in models
  that explicitly include the O orbitals, but also that our solution
  contradicts the long-held belief that the quantum spin fluctuations
  of the antiferromagnetic (AFM) background play a key role in
  determining this dynamics. Indeed, we show that the correct,
  experimentally observed dispersion is generically obtained for a
  hole moving on the O sublattice, and coupled to a N\'eel lattice of
  spins without spin fluctuations.  This marks a significant
  conceptual change in our understanding of the relevant phenomenology
  and opens the way to studying few-holes dynamics without finite-size
  effect issues \cite{BayoB}, to understand the actual strength of the
  ``magnetic glue''. }

The  simplification from three-band to one-band models is based on
the idea that the quasiparticle resulting when one hole is doped in
the system has predominantly Zhang-Rice singlet (ZRS) character
\cite{ZR,SE}. Agreement between the quasiparticle dispersion for a
generalized $tJ$ model (with longer range hopping) and that
measured by angle-resolved photoemission spectroscopy (ARPES) in
parent compounds \cite{ARPES,OK2,Wells,Andrea,Ronning,OK1} is taken as evidence that 
one-band models are valid. Whether this is a good approximation in all
the Brillouin zone and also for finite doping, or whether it is valid
only near the $({\pi\over2}, {\pi\over 2})$ minimum, is still debated
\cite{Bayo}. In one-band models, moreover, spin- and
charge-fluctuations arise from the same particle-hole excitations,
making it difficult to envisage a separation between the
quasiparticles and the pairing glue. Such a separation, however, is
assumed in most theories describing spin-fluctuations mediated pairing
\cite{rev1}.  Even more problematic are recent arguments that such a strong
attractive interaction mediated by spin-fluctuations is actually
ignored by one-band models \cite{Mirko}.  In other words, even if
these models capture the quasiparticle dispersion accurately, they may
still fail to properly describe their effective interactions.

To fully answer these questions, one needs to be able to
compare predictions of the three-band and one-band models not just in
the single hole sector, where a single quasiparticle forms and its
dispersion can be calculated, but also in the two-hole sector, where the
effective interactions between quasiparticles can be studied. Carrying
out two-hole calculations by exact numerical means is still too
difficult a task: quantum Monte Carlo algorithms suffer from sign
problems, while at present exact diagonalization (ED) can be carried
out only on rather small clusters, where the finite size effects are
still considerable and render the interpretation of the results
difficult \cite{BayoB}.

In this Article we show that a simple variational approximation for a
three-band model on an infinite lattice captures all main known
aspects of the quasiparticle behavior not just qualitatively, but also
quantitatively.  This approximation can also be systematically
improved by increasing the variational space; this provides an
estimate for the relevance of the excluded states. Most importantly,
this method can be straightforwardly generalized to calculate
few-hole propagators \cite{Mirko,MB}.

Here we present the one-hole solution which already reveals several
major surprises: (i) we find that the spin fluctuations of the AFM
background play a negligible role in determining the quasiparticle
dispersion, because the hole moves on a different sublattice. By
contrast, in one-band models it is widely believed that the dynamics
of a ZRS is controlled by these fluctuations, because not only does
the ZRS move in the magnetic sublattice but it is also a coherent mix
of spin and charge degrees of freedom. We argue that this view is
wrong, and that the necessary inclusion of longer-range hopping in
one-band models has precisely the effect of minimizing the role of the
spin fluctuations; (ii) the quasiparticle's dispersion in our model has the
characteristic shape measured experimentally for any reasonable choice
of parameters, unlike in one-band models where addition of longer
range hoppings is necessary to obtain the correct dispersion, as
mentioned above; (iii) our method allows us to study five-band models
to understand the importance of the in-plane O $2p$ orbitals which do
not hybridize directly with Cu $3d_{x^2-y^2}$. While, as expected, we
find that the quasiparticle dispersion is little affected, the ARPES
spectral weight is significantly changed and now exhibits a strong
suppression outside the magnetic Brillouin zone in agreement with
ARPES findings \cite{ARPES,Andrea}. This suggests that even three-band
models do not  fully capture all the quasiparticle properties.

\begin{figure}[t]
\includegraphics[width=\columnwidth]{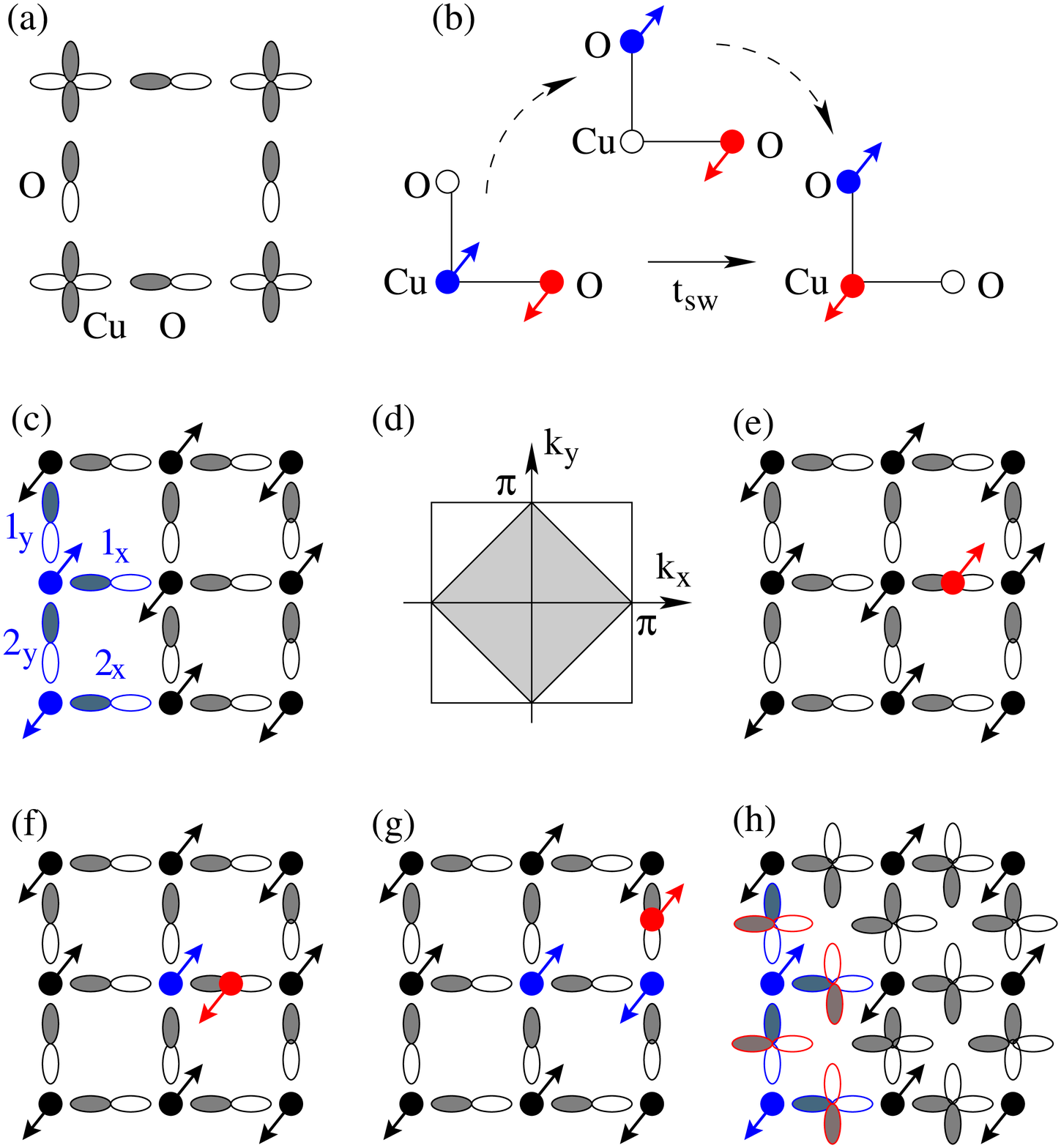}
\caption{(a) Sketch of three-band model which includes the Cu
  $3d_{x^2-y^2}$ and the O ligand $2p_{x/y}$ orbitals.  White/shaded
  areas indicate our choice for the positive/negative lobs; (b) Sketch
  of a spin-swap process which results in effective hopping of the hole
  while its spin is swapped with that of the neighbour Cu; (c) Unit cell
  for N\'eel AFM order, consisting of two Cu spins and four O orbitals
  (highlighted in blue); (d) Magnetic Brillouin zone (shaded area) and
  the full Brillouin zone; (e) Example of  a zero-magnon configuration,
  with the spin-up hole on an O orbital (red arrow) and the Cu
  spins with N\'eel order; (f) Example
  of a one-magnon configuration, where a Cu spin (marked in blue) is
  flipped, as is the hole's spin (red arrow); (g) Example of a
  two-magnon configuration, with two flipped Cu spins (shown in
  blue). The hole is spin-up again (red arrow); (h) Unit cell for
  N\'eel AFM order in the 
  five-band model. The additional O orbitals are highlighted in red.}
\label{fig1}
\end{figure}

The model we study can be thought of as the $tJ$ analog
of the three-band Emery model \cite{Emery}: double-occupancy on
the Cu sites is forbidden because of the large on-site Hubbard
repulsion, so there is a spin-${1\over 2}$ at each Cu site while the
doping hole enters the O $2p$ ligand orbitals, see
Figs. \ref{fig1}(a),(e). The  resulting Hamiltonian is \cite{Bayo}:
\begin{equation}
\EqLabel{e1}
{\cal H} = T_{pp} + T_{swap} + H_{J_{pd}} +H_{J_{dd}}.
\end{equation}
 $T_{pp}$ describes first and second nearest neighbour (nn) hopping
of the hole; $T_{swap}$ describes effective hopping of the hole
mediated by the Cu spin, whereby first the Cu hole hops onto a
neighbour O followed then by the original hole filling the
Cu
orbital, see Fig. \ref{fig1}(b). Note that this leads to
a swap of the spins of the hole and the Cu; $H_{J_{pd}}$ describes the
AFM exchange between 
the spins of the hole and of its two neighbour Cu; and
$H_{J_{dd}}$ describes the nn AFM superexchange between Cu spins
except on the bond occupied by the  hole. If 
$J_{dd}=1$ is the energy unit, then $t_{pp} = 4.13,
t_{pp}'=2.40, t_{sw}=2.98$ and $J_{pd}=2.83$, respectively. The reader
is referred to Ref.  \cite{Bayo} for further details on the
Hamiltonian, and on its ED solution for a hole on a 32 Cu $+$ 64 O
cluster. 

In order to study this Hamiltonian on an infinite lattice, we make the
key simplification of reducing $H_{J_{dd}}$ to an Ising form, instead
of its full Heisenberg form. As a result, the undoped ground-state
$|{\rm AFM}\rangle$ is a simple N\'eel state without any
spin-fluctuations. This approach will be justified {\it a posteriori}
based on the results it leads to.

The unit cell of the N\'eel AFM has two Cu spins and thus four distinct O
sites; this and the corresponding magnetic Brillouin zone (MBZ) are
shown in Figs. \ref{fig1}(c),(d). Thus, there are four
inequivalent hole Bloch states $p^\dagger_{{\bf k}, \alpha, \sigma} =
{1\over \sqrt{N}} \sum_{i\in A_{\alpha}}^{} e^{i{\bf k}
  \bf{R}_{i,\alpha}} p^\dagger_{i,\alpha,\sigma}$, where $N\rightarrow
\infty$ is the number of unit cells, $\alpha\in \{1_x,1_y,2_x,2_y\}$
labels the type of O orbital while $A_{\alpha}$ is the sublattice of all O of
type $\alpha$, ${\bf R}_{i,\alpha}$ is the location of the $\alpha$ O
of unit cell $i$, ${\bf k}$ is a quasi-momentum inside the MBZ and
$p^\dagger_{i,\alpha,\sigma}$ creates a spin-$\sigma$ hole at O$_{i,\alpha}$. In
the following we set $\sigma=\uparrow$ (the $\sigma=\downarrow$ case
is treated similarly and gives identical results) and define the
single-hole propagators:
\begin{equation}
\EqLabel{e2}
G_{\beta\alpha}({\bf k}, \omega) = \langle {\rm AFM}| p_{{\bf
    k},\beta,\uparrow} \hat{G}(\omega) p^\dagger_{{\bf k},
  \alpha,\uparrow} |{\rm AFM}\rangle
\end{equation}
where $ \hat{G}(\omega)= [\omega + i \eta - {\cal H}]^{-1}$, $\hbar=1$
and $\eta> 0$ is a small 
broadening. The energy $\omega$ is measured from the undoped
ground-state, i.e. we set ${\cal H}_{J_{dd}}|{\rm
  AFM}\rangle=0$. The one-hole spectrum $E_n({\bf k})$ is given by the poles of 
these propagators, while from the  residues one can find the
overlaps $\langle n, {\bf k},\uparrow|p^\dagger_{{\bf k},
  \alpha,\uparrow} |{\rm AFM}\rangle$, where ${\cal H}| n, {\bf
  k},\uparrow\rangle= E_n({\bf k}) | n, {\bf k},\uparrow\rangle$ are
the one-hole eigenstates for band $n$.

To calculate these propagators, we use the identity $\hat{G}(\omega)
(\omega + i\eta - {\cal H})=1$ to find $(\omega+i\eta)
G_{\beta\alpha}({\bf k}, \omega) = \delta_{\alpha,\beta} + \langle
{\rm AFM}| p_{{\bf k},\beta,\uparrow} \hat{G}(\omega) {\cal H}
p^\dagger_{{\bf k}, \alpha,\uparrow} |{\rm AFM}\rangle$. The
Hamiltonian has (i) terms which do not change either the hole location
or its spin (${\cal H}_{J_{dd}}$ and the diagonal part of ${\cal
  H}_{J_{pd}}$) and lead to a simple energy shift; (ii) terms which
change the hole location but not its spin ($T_{pp}$ and terms in
$T_{swap}$ which move the hole past the Cu with the same spin
orientation) and link $G_{\beta\alpha}$ to other $G_{\beta\alpha'}$;
and (iii) terms which flip the hole's spin, while also flipping a
neighbouring Cu spin (terms in $T_{swap}$ which move the hole past the
Cu with antiparallel spin, and the off-diagonal part of ${\cal
  H}_{J_{pd}}$). These last terms define generalized propagators which
we call one-magnon propagators because they are projected on states
that have a magnon (flipped Cu spin) beside the hole. One example of
such a state is shown in Fig. \ref{fig1}(f). Equations of motion for
the one-magnon propagators are obtained similarly, and link them to
other one-magnon propagators with a different hole-magnon distance, to
two-magnon propagators like shown in Fig. \ref{fig1}(g), since the
hole can flip a second Cu spin, and -- if the hole and magnon are on
neighbouring sites -- back to various $G_{\beta \alpha}$. The
equations for two-magnon propagators link them to other two- and
three- , and possibly also to one-magnon propagators, and so on and so
forth. While the full set of exact equations of motion can be thus
generated, they are impossible to solve exactly.

We introduce a variational solution using the fact that each time a
new magnon is created, the energy is increased by (about) $2 J_{dd}$
since up to four Cu-Cu bonds become FM. Many-magnon states are thus
energetically expensive and unlikely to be significant components of
the lowest-energy eigenstates. We define a variational approximation
by choosing an integer $n_{m}$ and setting all propagators with more
than $n_{m}$ magnons to zero. This leads to a manageable (although
still infinite) sparse system of equations that can be solved
efficiently.

\begin{figure}[t]
\includegraphics[width=\columnwidth]{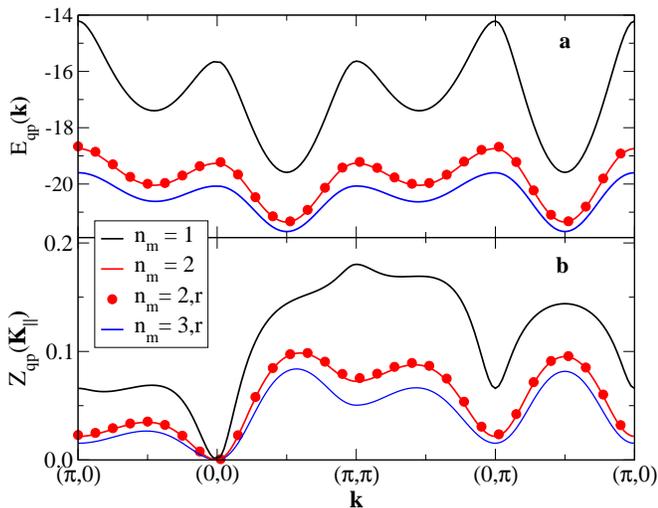}
\caption{(a) Quasiparticle dispersion (in units of
  $J_{dd}$)  along various
  cuts in the BZ. The results are for the three-band model using
  $n_m=1$ (black line),  $n_m=2$ (red line), restricted
  $n_m=2$ (red circles) and restricted $n_m=3$ (blue line)
  approximations. In the restricted approximations, only 
  configurations with magnons on adjacent sites are included. (b) The
  corresponding ARPES quasiparticle spectral weights.} 
\label{fig2}
\end{figure}

Since ED results show a distortion of the AFM background only rather
close to the hole (Fig. 3 of Ref. \cite{Bayo}) it is reasonable to
expect that small $n_{m}$ may already give a good approximation. To
check this, we  calculate the results for $n_m\le 3$. For
  $n_m=2$ we do both the full variational calculation that allows the
  magnons to be at any distance from one another, and the restricted
  calculation where only configurations with the two magnons on
  adjacent sites are kept (the hole can be located anywhere).  In the
  $n_m=3$ case we perform only the restricted calculation where
  the magnons are in a connected cluster.  The corresponding
dispersions of the low-energy quasiparticle are shown in
Fig. \ref{fig2}(a) along several cuts in the full Brillouin zone
(FBZ).

The most striking observation is that the dispersions have a shape
similar to that measured experimentally, with deep isotropic minima at $\left(
{\pi\over2},{\pi\over2}\right)$. This shows that even the very simple
$n_m=1$ solution already captures important aspects of the correct
quasiparticle dynamics.

As expected for bigger variational spaces, the dispersions for larger
$n_m$ lie at lower energies.  The bandwidths for $n_m=2,3$ are about
half of that for $n_m=1$, due to standard polaronic physics. Consider
$n_m=2$: while it is energetically favourable for the hole to be near
the secondly emitted magnon, as they have antiparallel spins,
configurations with the hole near the first magnon are not favorable
because of their parallel spins. If the first magnon is bound in the
cloud it is in configurations like in Fig. \ref{fig1}(g), where its
location limits the number of broken AFM bonds. Alternatively, this
magnon can dissociate from the cloud resulting in excited states
starting from $E_{\rm 1,gs} + 2 J_{dd}$, i.e. the ground-state energy
of the $n_m=1$ quasiparticle plus the $2J_{dd}$ cost for a magnon
located far from it. For our parameters, this continuum starts at
$\approx -17.58 J_{dd}$ so the $n_m=2$ quasiparticle band must become narrower
in order to fit below it. The comparison between the full and the
restricted $n_m=2$ cases confirms that the connected magnon clusters
(which cost less exchange energy) account for the overwhelming
contribution to the low-energy quasiparticle, as expected.

The $n_m=3,r$ results show an additional narrowing of the
 bandwidth from $2.6J_{dd}$ for $n_m=2$, to $2.05
J_{dd}$. This solution is thus very close to the $2 J_{dd}$ bandwidth
of the fully converged case. This is not surprising since the
quasiparticle cannot possibly bind too many magnons in its cloud,
given that each magnon is at a different location and that the hole
can interact with at most one favorable magnon (with antiparallel
spin) in any configuration. We conclude that the $n_m=3,r$ solution is
already quantitatively accurate, and indeed its dispersion is in
excellent agreement with the ED dispersion of Ref. \cite{Bayo}.

This quantitative agreement between the variational and ED results
shows that quantum spin fluctuations of the AFM background (fully
included in ED but frozen in our variational approach) have little or
no effect on the quasiparticle's dynamics.  This is because in
three-band models the hole can move freely on the O sublattice, so it can 
easily go to absorb magnons created previously and then emit others at new locations to move the cloud, resulting in
fast quasiparticle dynamics. Spin fluctuations of the background,
which act on a slower time scale ($J_{dd}$ is the smallest energy) are
then not essential for this dynamics. Indeed, we have attempted
to gauge the effect of spin fluctuations for $n_m=2$ by adding in the
equations of motion terms that directly link two-magnon and $G_{\beta
  \alpha}$ propagators, mimicking spin fluctuations that either
produce or remove a pair of nn magnons close to the hole. Such terms
lead to very minor quantitative changes, as will be reported elsewhere
\cite{Hadi}.

 This conclusion may seem surprising since for one-band models it is
 believed that spin fluctuations are essential in determining the
 quasiparticle dispersion: as a ZRS moves it creates a string of
 wrongly oriented spins (magnons) whose energy increases linearly with
 its length, and which ``ties'' it near the starting position. In the
 absence of spin fluctuations, the quasiparticle acquires a finite
 mass only by executing Trugman loops \cite{Trugman} which are
 many-step (and thus very slow) processes that lead to a very heavy
 quasiparticle \cite{Mhol}. Spin fluctuations act faster to remove
 pairs of nn magnons from the string and thus release the ZRS. These
 arguments, however, depend essentially on the assumption that only nn
 hopping of the ZRS is possible, despite the knowledge that the
 resulting dispersion is wrong, being nearly flat along
 $(0,\pi)-(\pi,0)$. To obtain agreement with experiments, second and
 third nn hopping must be added \cite{ARPES, Wells}. These allow the
 ZRS to move freely on its magnetic sublattice and get away from the
 string of defects that nn hopping creates, similar to what happens in
 three-band models. The longer-range hopping thus changes the
 phenomenology qualitatively and in its presence, we find that
 spin fluctuations are no longer
 essential for the quasiparticle dynamics in one-band models either,
 unlike when only nn hopping 
 is allowed \cite{Hadi}.

\begin{figure}[t]
\includegraphics[width=\columnwidth]{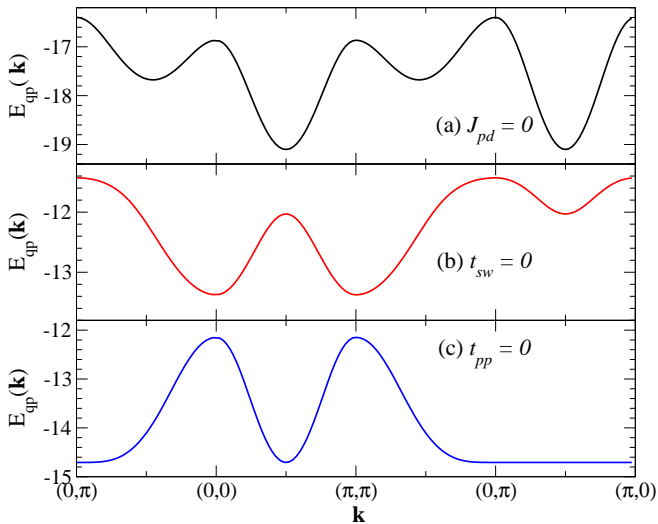}
\caption{Quasiparticle dispersion for
  the three-band model and $n_m=2$, when we set (a) $J_{pd}=0$; (b)
  $t_{sw}=0$; (c) $t_{pp}=0$. The other parameters are kept at
  their stated values. }
\label{fig3}
\end{figure}

A natural follow-up question is whether careful tuning of the
parameters is needed to achieve this dispersion, or whether this
shape is generic. The answer is the latter. Specifically, ${\cal
  H}_{J_{pd}}$ has almost no effect on the shape of $E_{qp}({\bf k})$:
even setting $J_{pd}=0$ leaves it virtually unchanged, only shifting
the overall value as exchange energy is lost, see Fig. \ref{fig3}(a).
Setting either $t_{sw}=0$ or $t_{pp}=0$ leads to very different
dispersions (Figs. \ref{fig3}(b), (c)), however if $t_{pp}$ and
$t_{sw}$ are comparable, the correct shape appears. In fact, a deep
minimum at $\left( {\pi\over2},{\pi\over2}\right)$ is then achieved
even for $n_m=0$ (not shown). This confirms the speculation in
Ref. \cite{Bayo} that $E_{qp}({\bf k})$ arises through constructive
interference between $T_{pp}$ and $T_{swap}$, and shows that both
terms are needed to properly describe the quasiparticle dynamics. Note
that many studies of three-band models ignore $T_{pp}$ or treat it as
a perturbation \cite{mod1,mod2,mod3, mod4} (for more discussion, see
the supplementary material of Ref. \cite{Bayo}).

Having established that the shape of $E_{qp}({\bf k})$ is robust, we
now analyze the quasiparticle ARPES weight. If ${\bf K}=({\bf
  K}_{\parallel}, K_z)$ is the 
photoelectron's momentum and $\omega$ is the transferred energy,
and assuming an unpolarized beam, 
 the ARPES intensity \cite{Mau} is $A({\bf K},
\omega) \sim \sum_{{\bf k}, {\bf 
    G}}^{} \delta_{{\bf K}_{\parallel}+ {\bf k}, {\bf G}}
\sum_{\alpha,\beta}^{} e^{i{\bf G}\cdot{\bf R}_{\alpha\beta}}
\eta_{\alpha\beta} A_{\alpha\beta}({\bf k},\omega)$. We checked
that this gives the correct unfolding if we decouple the hole from
the spins, since then the dispersion in the FBZ can be calculated analytically.
  Here ${\bf G}$
are the reciprocal lattice vectors of the MBZ and ${\bf k}$ are
momenta in the first MBZ. The first sum shows
that ARPES detects quasiparticles of quasi-momentum ${\bf k}$ equal to
the photohole's in-plane momentum $-{\bf K}_{\parallel}$, modulo ${\bf
  G}$. ${\bf R}_{\alpha \beta} = {\bf R}_{i,\alpha}-{\bf
  R}_{i,\beta}$ is the distance between the O sites $\alpha, \beta$
and $A_{\alpha\beta}({\bf k},\omega)=-{1\over \pi} \mbox{Im}
G_{\alpha\beta}({\bf k},\omega)$ are the spectral weights of the
sublattice propagators. Finally, $\eta_{\alpha\beta}=1$ if the
orbitals $\alpha$ and $\beta$ are both either $2p_x$ or $2p_y$, and
zero otherwise. The quasiparticle ARPES spectral weight,  $Z_{qp}({\bf
  K}_{\parallel})$, is the weight at the energy $\omega = E_{qp}({\bf k})$ of the
quasiparticle, {\em i.e.} $A({\bf K}, \omega\rightarrow E_{qp}({\bf k}))
\rightarrow Z_{qp}({\bf 
  K}_{\parallel}) \delta(\omega - E_{qp}({\bf k}))$.

In Fig. \ref{fig2}(b) we plot $Z_{qp}({\bf K}_{\parallel})$ along
various cuts in the FBZ. The first observation is that unlike $E_{qp}({\bf k})$,
$Z_{qp}({\bf K}_{\parallel})$ does not have MBZ periodicity: the evolution
along  $(0,0)-(\pi,\pi)$ is not symmetric about $\left( {\pi\over2},
{\pi\over2}\right)$. This is expected: while
 all $A_{\alpha\beta}({\bf k},\omega)$ must, and indeed do,
exhibit MBZ periodicity, $A({\bf K}, \omega)$ does not because of the
$e^{i{\bf G}\cdot{\bf R}_{\alpha\beta}}$ phases. If ${\bf K}_{\parallel}$
is inside the first MBZ then ${\bf G}=0$ and, for example, ARPES
measures constructive interference between the two $2p_x$ orbitals'
contributions. If ${\bf K}_{\parallel}$ crosses  into the
second MBZ, then ${\bf G}= (\pm\pi,\pm\pi)$ and ARPES measures
destructive interference between these two orbitals since they are at
a distance
$(0, a=1)$ apart. 

It is worth pointing out that a similar approach (N\'eel order plus a
few magnons) for one-band models does not lead to any asymmetry. This
is because even though there are two sublattice Bloch-states with the
ZRS located on either magnetic sublattice, there is no interference
between them as they belong to sectors with different total spin
$S_z$. Additional Hubbard and spin-fluctuations corrections must be
included to obtain an asymmetric spectral weight, see for example
Ref. \cite{Sushkov}.

The second observation is that $Z_{qp}({\bf K}_{\parallel})$ disagrees
along the $(0,0)-(\pi,\pi)$ cut  with the experimental
measurements which find large weight near $\left({\pi\over2},
{\pi\over 2}\right)$ that decreases fast on both sides
\cite{ARPES,Andrea}. (ED predicts 
$Z_{qp}(\pi,\pi)=0$ because its quasiparticle has spin 3/2 in that
region. Such an object cannot be fully described with a N\'eel
background which breaks invariance to spin rotations). Since the situation 
improves with increasing $n_m$, it is possible that
 going to higher $n_m$ may fix
this problem. However, such an explanation is rather unsatisfactory
because it suggests a sensitive dependence of the ARPES  weight on the
precise structure of the magnon cloud, unlike the robust insensitivity
of the dispersion.

To check for an alternative explanation, we  add the second set of
in-plane O $2p$ orbitals to our 
model, resulting in the new unit cell sketched in
Fig. \ref{fig1}(h). These orbitals are usually ignored because they do not
hybridize directly with the Cu $3d_{x^2-y^2}$ orbitals. However, they do
hybridize  heavily with the ligand $2p$  orbitals occupied by the
hole, so their role should be evaluated more carefully and this can be
done easily with our method.

For the Hamiltonian, this requires us to expand $T_{pp}$
accordingly. This is achieved without introducing new parameters because nn
hopping between two new orbitals also has  magnitude $t_{pp}$,
while between new and old orbitals  $\tilde{t}_{pp}/t_{pp}=
(t_{pp,\sigma} - t_{pp,\pi})/(t_{pp,\sigma} + t_{pp,\pi})=0.6$ since
$t_{pp,\pi}=t_{pp,\sigma}/4$ . We can also add nnn hopping
$\tilde{t}_{pp}'$ for the new orbitals. Since $t_{pp,\sigma}$ scales with distance like
$1/d^4$, it follows that $\tilde{t}_{pp}' = t_{pp,\sigma}/4 = 0.2
t_{pp}$ \cite{Harisson}. This is  smaller than
$t_{pp}'\approx 0.6 t_{pp}$ for the old orbitals for whom nnn
hopping is boosted through hybridization with the $4s$ orbital of the
bridging Cu. In any event, we find very little sensitivity to the
precise values
we use for $\tilde{t}_{pp}'$  \cite{Hadi}.

\begin{figure}[t]
\includegraphics[width=\columnwidth]{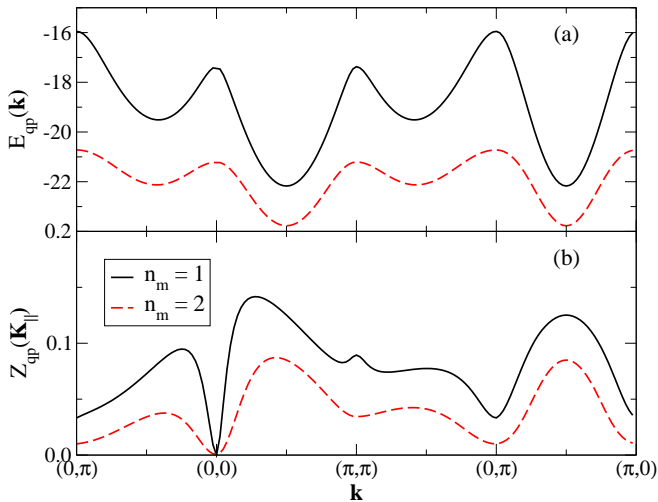}
\caption{Same as in Fig. \ref{fig2}, but for the five-band model.}
\label{fig4}
\end{figure}

We study the five-band model with the same variational approximations;
now there are 64 sublattice propagators $G_{\beta\alpha}({\bf k},
\omega)$, leading to a corresponding increase in the number of
equations of motion. Fig. \ref{fig4} shows the quasiparticle
dispersion and ARPES spectral weight for the $n_m=1,2$ solutions. For
$E_{qp}({\bf k})$, the results are very similar to the results shown
in Fig. \ref{fig2}(a), but the bands are somewhat wider, as expected
because of the increased bare kinetic energy. We have checked that the
dependence on $J_{pd}, t_{sw}$ and $t_{pp}$ is essentially
unchanged. Indeed, the expectation that this other set of orbitals has 
little effect on the quasiparticle dynamics is correct.

However, their addition has a significant effect on the
evolution of $Z_{qp}({\bf K}_{\parallel})$ on the $(0,0)-(\pi,\pi)$
cut. The asymmetry is maintained but the results now show a decrease
of the ARPES spectral weight on both sides of the MBZ boundary, in
 agreement with experimental data \cite{ARPES,Andrea}.

The fact that the weight is significantly changed for the
five-band model vs. the three-band model even though the dispersion is
not much affected should not be a surprise. Since ARPES measures
interference between like $2p$ orbitals, the quasiparticle weight can
be significantly affected even by rather small redistributions of the
wavefunction among orbitals, unlike the energy. These results
suggest that a full understanding of the evolution of the spectral
weight at low dopings, currently still missing, may require
inclusion into theoretical models of these additional
orbitals. This will only increase the need for
accurate approximations like the one we propose here, since exact
numerical approaches become even more challenging to implement  in
larger Hilbert spaces. 

To summarize, we used a simple variational approach to study a quasiparticle in three- and 
five-band models of an infinite CuO$_2$ layer, while also being able
to gauge  accuracy by increasing the
variational space. Our results compare well with available results from ED of
small clusters. 
Since the variational approach ignores the effect of spin-fluctuations
in the AFM layer, the good agreement for the dispersion strongly
supports the idea that these spin fluctuations do not play the
important role in the quasiparticle dynamics
attributed to them based on results for one-band models with only nn hopping.

This is a very important finding because properly describing the
background spin fluctuations  is very difficult and a
major barrier to studying the two-hole sector to understand the
effective interactions between quasiparticles, which is the second
piece of knowledge (besides the quasiparticle dispersion) needed in
order to propose accurate simple(r) effective models. Our method
allows us to distinguish the magnons emitted and absorbed by holes,
which are treated exactly, from those due to background fluctuations,
which are ignored. Since the method also generalizes to treat few-hole
states, we are now able to investigate the role of magnon exchange in
mediating strong attractions between holes, and to verify whether this
attraction is indeed absent from the currently used one-band effective models, as
speculated in Ref. \cite{Mirko}. This work is now in progress.

\acknowledgments
Acknowledgements: We  thank B. Lau and W. Metzner for insightful comments. This
work was funded by QMI, CIfAR 
and NSERC.

Competing interests statement: 
The authors declare no competing financial interests.

\end{document}